\documentclass[article,preprint,superscriptaddress,graphicx]{revtex4}
\usepackage[dvipdfmx]{graphicx} 
\usepackage{amsmath,amssymb}
\draft % marks overfull lines with a black rule on the right

\begin{document}

%\preprint{}

\title{Anomalous behavior of $1/f$ noise in graphene near the charge neutrality point} %Title of paper



\author{Shunpei Takeshita}
%\thanks{}
\affiliation{Graduate School of Science, Osaka University, Toyonaka, Osaka 560-0043, Japan}
\author{Sadashige Matsuo}
%\thanks{}
\affiliation{Graduate School of Science, Osaka University, Toyonaka, Osaka 560-0043, Japan}
\affiliation{Institute for Chemical Research, Kyoto University, Uji, Kyoto 611-0011, Japan}
\author{Takahiro Tanaka}
%\thanks{}
\affiliation{Graduate School of Science, Osaka University, Toyonaka, Osaka 560-0043, Japan}
\author{Shu Nakaharai}
%\thanks{}
\affiliation{WPI-MANA, NIMS, Tsukuba, Ibaraki 305-0044, Japan}
\author{Kazuhito Tsukagoshi}
%\thanks{}
\affiliation{WPI-MANA, NIMS, Tsukuba, Ibaraki 305-0044, Japan}
\author{Takahiro Moriyama}
%\thanks{}
\affiliation{Institute for Chemical Research, Kyoto University, Uji, Kyoto 611-0011, Japan}
\author{Teruo Ono}
%\thanks{}
\affiliation{Institute for Chemical Research, Kyoto University, Uji, Kyoto 611-0011, Japan}
\author{Tomonori Arakawa}
%\thanks{}
\affiliation{Graduate School of Science, Osaka University, Toyonaka, Osaka 560-0043, Japan}
\author{Kensuke Kobayashi}
%\thanks{}
\affiliation{Graduate School of Science, Osaka University, Toyonaka, Osaka 560-0043, Japan}


\date{\today}

\begin{abstract}
We investigate the noise in single layer graphene devices from equilibrium to far-from equilibrium and found that the $1/f$ noise shows an anomalous dependence on the source-drain bias voltage ($V_{\mathrm{SD}}$). While the Hooge's relation is not the case around the charge neutrality point, we found that it is recovered at very low $V_{\mathrm{SD}}$ region.  We propose that the depinning of the electron-hole puddles is induced at finite $V_{\mathrm{SD}}$, which may explain this anomalous noise behavior.
\end{abstract}


\maketitle %\maketitle must follow title, authors, abstract and \pacs
Graphene is a two dimensional material with gate-tunable high-mobility carriers, which makes it one of the most promising materials for new electronic devices such as high speed transistors~\cite{Graphenetransistor}. As the usefulness of the devices is hampered by electrical noise, to investigate the noise mechanism is significant~\cite{1/fnoisereview}.  In addition to the purpose for the applications, it is scientifically important to investigate the noise as it gives us a variety of information on the electronic transport.

The $1/f$ noise is the electrical noise observed in many electronic devices~\cite{1/fnoisereview}. Various physical mechanisms such as trapping-detrapping process of carriers or disorder scattering cause the resistance fluctuation and thus the $1/f$ noise. There is an empirical expression for the $1/f$ noise, namely, Hooge's relation~\cite{Hooge}, which widely holds  true in conventional metals and semiconductors. It tells that the voltage noise power spectral density $S_V$ is expressed by
\begin{equation}
S_V(f)=A\frac{V_{\mathrm{SD}}^2}{f}=\frac{\alpha_{\mathrm{H}}}{N}\frac{V_{\mathrm{SD}}^2}{f},
\label{Hooge}
\end{equation}
where $f$ and $V_{\mathrm{SD}}$ are frequency and source-drain bias voltage, respectively. Here $A$ is called noise amplitude. Usually, $A$ is inversely proportional to the number of charge carriers in the sample $N$, and is proportional to Hooge's parameter $\alpha_{\mathrm{H}}$. Although a rigorous theoretical background for this relation is not yet established, the reduction of noise amplitude with an increase of $N$ in principle occurs through a reduction in the relative fluctuation in the number of charge carrier or through a better screening of charged scattering centers.

So far, several groups reported that the noise amplitude of the $1/f$ noise in graphene shows non-monotonic gate voltage dependence~\cite{M1,M2,M3,M4,M5,M6,NM1,NM2,NM3} unlike a naive expectation from Hooge's relation; the noise amplitude unexpectedly becomes minimum around the Dirac point, or the charge neutrality point (CNP), where the carrier density is minimum. As the distance from the CNP increases by changing the gate voltage, the noise amplitude increases to its local maximum and then decreases again, leading to the ``M-shape'' gate voltage dependence~\cite{M1,M2,M3,M4,M5,M6}. 

Here, we report an anomalous dependence on the source-drain bias voltage of the $1/f$ noise in single layer graphene devices. The previous experiments were conducted in the large $V_{\mathrm{SD}}$ region, which is several tens of mV~\cite{M1,M3,M4,M5,M6,NM1,NM2,NM3}. In contrast, we systematically studied the $1/f$ noise for wide $V_{\mathrm{SD}}$ region and observed that the behavior of the noise amplitude is strongly dependent on $V_{\mathrm{SD}}$; at high $V_{\mathrm{SD}}$ region, the ``M-shape'' gate voltage dependence is observed, while Hooge's relation is recovered at very low $V_{\mathrm{SD}}$ region. We propose that the $V_{\mathrm{SD}}$-induced depinning of the electron-hole puddles is likely to explain this anomalous noise behavior. 

\begin{figure}[tbp]
\begin{center}
\includegraphics[width=1\linewidth]{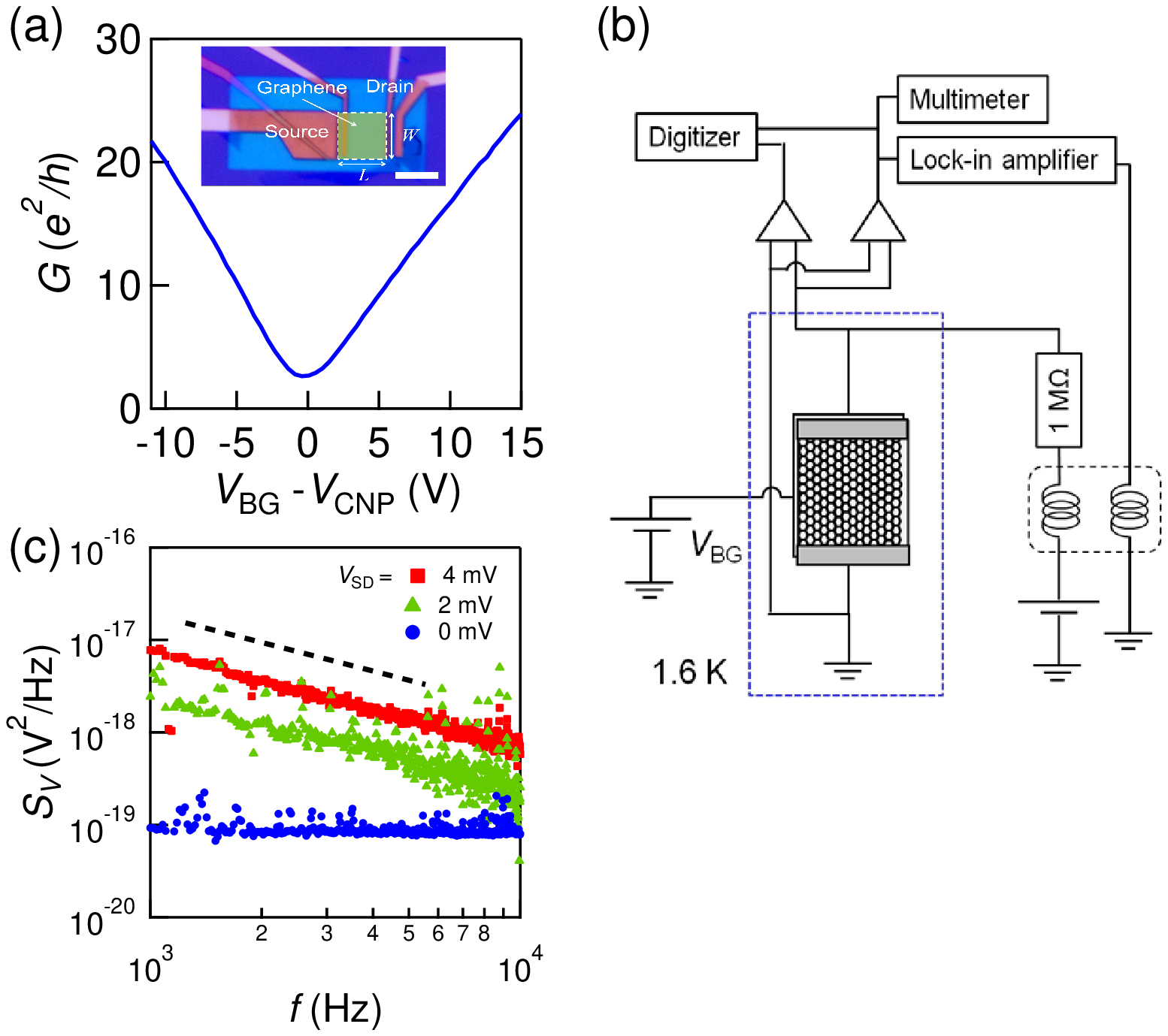}
\end{center}
\caption{(a) Conductance $G$ as a function of $V_{\mathrm{BG}}$. Inset shows the optical picture of the device  $\sharp1$. The scale bar is $10\,\mathrm{\mu m}$. The graphene is colered by the shaded square. The length ($L$) and the width ($W$) of the device are $L \sim 11.0\,\mathrm{\mu m}$ and $W \sim 12.7\,\mathrm{\mu m}$ respectively. The source and drain electrodes are formed by Pd/Au (5 /30 nm). (b) Schematic configuration of the measurement set up. The graphene device is at $1.6\,\mathrm{K}$ in the helium gas flow. (c) Typical voltage noise spectra at $V_{\mathrm{SD}}=0\,\mathrm{mV}$ (circle), $V_{\mathrm{SD}}=2\,\mathrm{mV}$ (triangle) and $V_{\mathrm{SD}}=4\,\mathrm{mV}$ (square). The dashed line indicates the $1/f$ dependence. }
\label{Fig1}
\end{figure}

A single layer graphene film was exfoliated on a $\mathrm{Si/SiO_2}$ substrate using the conventional mechanical exfoliation method~\cite{Electricfield}. The $\mathrm{Si/SiO_2}$ substrate works as a back gate to tune the carrier density in graphene by applying the voltage ($V_{\mathrm{BG}}$). The optical picture of the device is shown in the inset of Fig.~\ref{Fig1}(a).  The conductance and the noise measurements were performed at $1.6\,\mathrm{K}$ in the variable temperature insert (VTI by Oxford Inc.) as shown in the schematic picture of the measurement setup in Fig.~\ref{Fig1}(b)~\cite{Arakawa, Matsuo}.  The dc current with a small ac modulation was applied to the graphene sample through a $1\,\mathrm{M\Omega}$ resistor to obtain the differential conductance $G$ via standard lock-in technique. To measure the voltage noise spectral density at the device, two pairs of lead were connected across the device and the two voltage signals are independently amplified by the two room-temperature amplifiers (LI-75A by NF Corporation). In order to reduce the external noise, the measured two sets of the time domain data were cross-correlated to yield the noise power spectral density through the fast Fourier transformation (FFT). We measured two graphene devices ($\sharp1$ and $\sharp2$) with almost same size and geometry and obtained the consistent result regarding to the $V_{\mathrm{SD}}$-dependent behavior of the noise amplitude. Here we discuss the result obtained in the device $\sharp1$.

The gate voltage dependence of the conductance at zero source-drain bias voltage ($V_{\mathrm{SD}}=0\,\mathrm{V}$) is shown in Fig.~\ref{Fig1}(a) as a function of $V_{\mathrm{BG}} -V_{\mathrm{CNP}}$. Here $V_{\mathrm{CNP}}=4.0\,\mathrm{V}$ is the back gate voltage which gives the conductance minimum, signaling the CNP. The electron and hole mobilities were estimated from Fig.~\ref{Fig1}(a) to be $\mu_e=4\times10^3\,\mathrm{cm^2/Vs}$ and $\mu_h=5\times10^3\,\mathrm{cm^2/Vs}$, respectively~\cite{Matsuo}. These mobilities were not so high, which may be due to the PMMA resist covering the graphene.

Fig.~\ref{Fig1}(c) shows a typical measured noise spectral density as a function of frequency. At $V_{\mathrm{SD}}=0\,\mathrm{V}$, the noise, which consists of the thermal noise of the device and the residual amplifier noise, is independent of $f$. At a finite $V_{\mathrm{SD}}$, on the other hand, the noise becomes dependent on $f$ such that it is proportional to $1/f^{\beta}$, where $\beta\sim1$, after the frequency-independent noises are subtracted. At each $V_{\mathrm{BG}}$, the noise is measured for $-4\,\mathrm{mV}\leq V_{\mathrm{SD}}\leq 4 \,\mathrm{mV}$.

\begin{figure}[tbp]
\begin{center}
\includegraphics[width=0.9\linewidth]{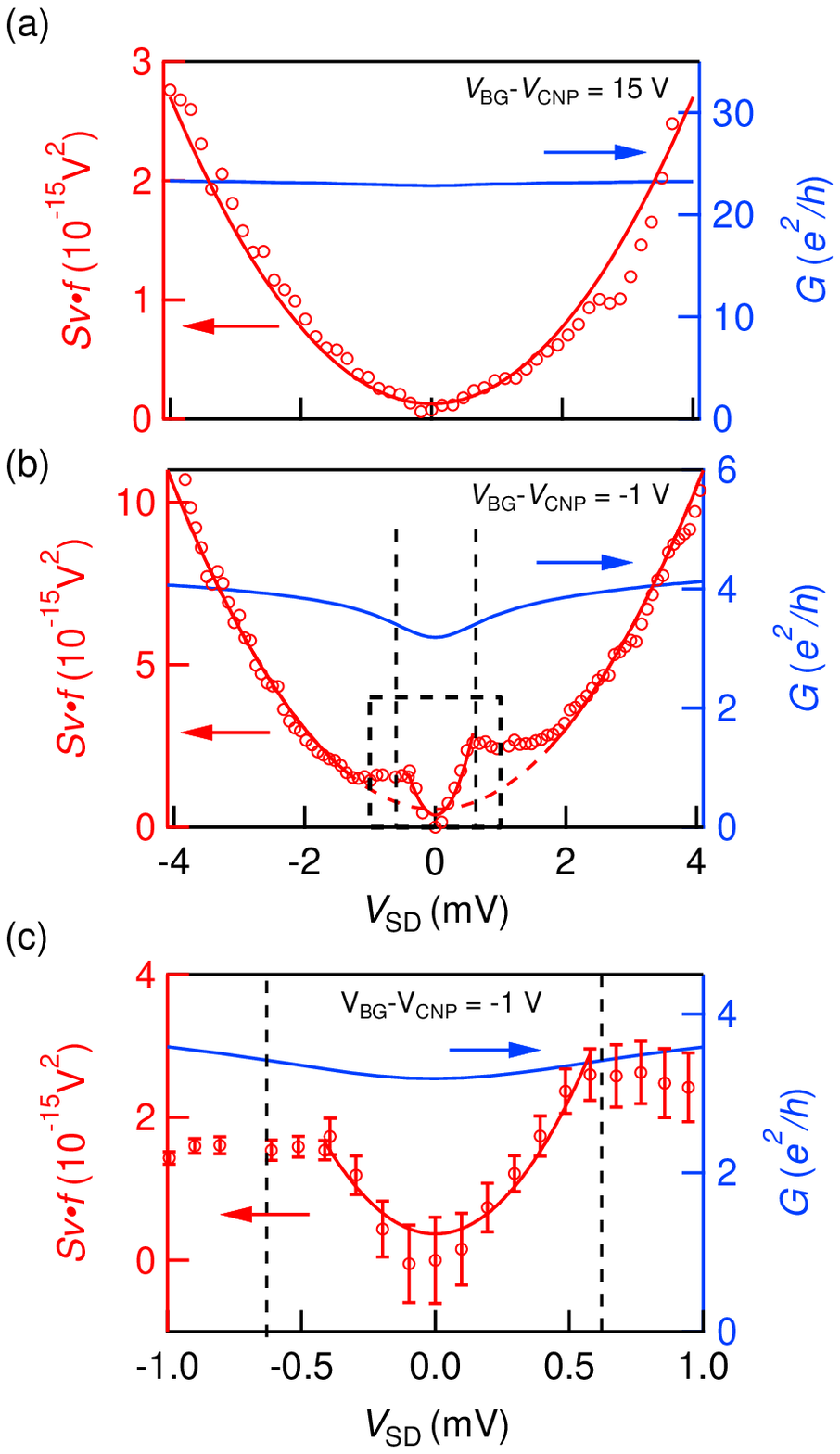}
\end{center}
\caption{Frequency normalized noise $S_V\cdot$$f$ (left axis) and the differential conductance at finite $V_\mathrm{SD}$ (right axis) as a function of $V_{\mathrm{SD}}$.  The red solid curves are the results of the numerical fitting to the relation $S_V\cdot$$f=AV_{\mathrm{SD}}^2$. (a) away from the CNP ($V_{\mathrm{BG}}-V_{\mathrm{CNP}}=15\,\mathrm{V}$) and (b) near the CNP ($V_{\mathrm{BG}}-V_{\mathrm{CNP}}=-1\,\mathrm{V}$). (c) Expanded view of the region surrounded by the bold dashed line in (b). The fine dashed lines in (b) and (c) indicate to $\pm V_{\mathrm{SD}}^\mathrm{C}$. 
}
\label{Fig2}
\end{figure}

From the obtained noise in the range between $1\,\mathrm{kHz}$ and $10\,\mathrm{kHz}$, the numerical fitting is performed to deduce the frequency normalized noise $S_V\cdot$$f$. The obtained $S_V\cdot$$f$ is plotted as a function of $V_{\mathrm{SD}}$ with the differential conductance at finite $V_\mathrm{SD}$ away from the CNP ($V_{\mathrm{BG}}-V_{\mathrm{CNP}}=15\,\mathrm{V}$) in Fig.~\ref{Fig2}(a) and near the CNP ($V_{\mathrm{BG}}-V_{\mathrm{CNP}}=-1\,\mathrm{V}$) in Fig.~\ref{Fig2}(b). The $S_V\cdot$$f$ away from the CNP shows an expected parabolic dependence to $V_{\mathrm{SD}}$. This confirms that the noise arises from the resistance fluctuation, which is converted into the voltage noise by the current. Although  $S_V\cdot$$f$ near the CNP also shows a parabolic dependence, the shapes changes around $|V_{\mathrm{SD}}|=V_{\mathrm{SD}}^\mathrm{C}\sim0.6\,\mathrm{mV}$ as shown in Figs.~\ref{Fig2}(b) and (c). Figure~\ref{Fig2}(c) is an expanded view of the region surrounded by the bold dashed line in Fig.~\ref{Fig2}(b). At the same time, the conductance shows a slight $V_{\mathrm{SD}}$-dependence in the regime of $|V_{\mathrm{SD}}| \leq V_{\mathrm{SD}}^\mathrm{C}$ as shown in Fig. 2(b). These may come from the same mechanism as we discuss below. The similar behaviors of the $S_V\cdot$$f$ around $\pm V_{\mathrm{SD}}^\mathrm{C}$ were observed for other gate voltages near the CNP.

\begin{figure}[tbp]
\begin{center}
\includegraphics[width=0.9\linewidth]{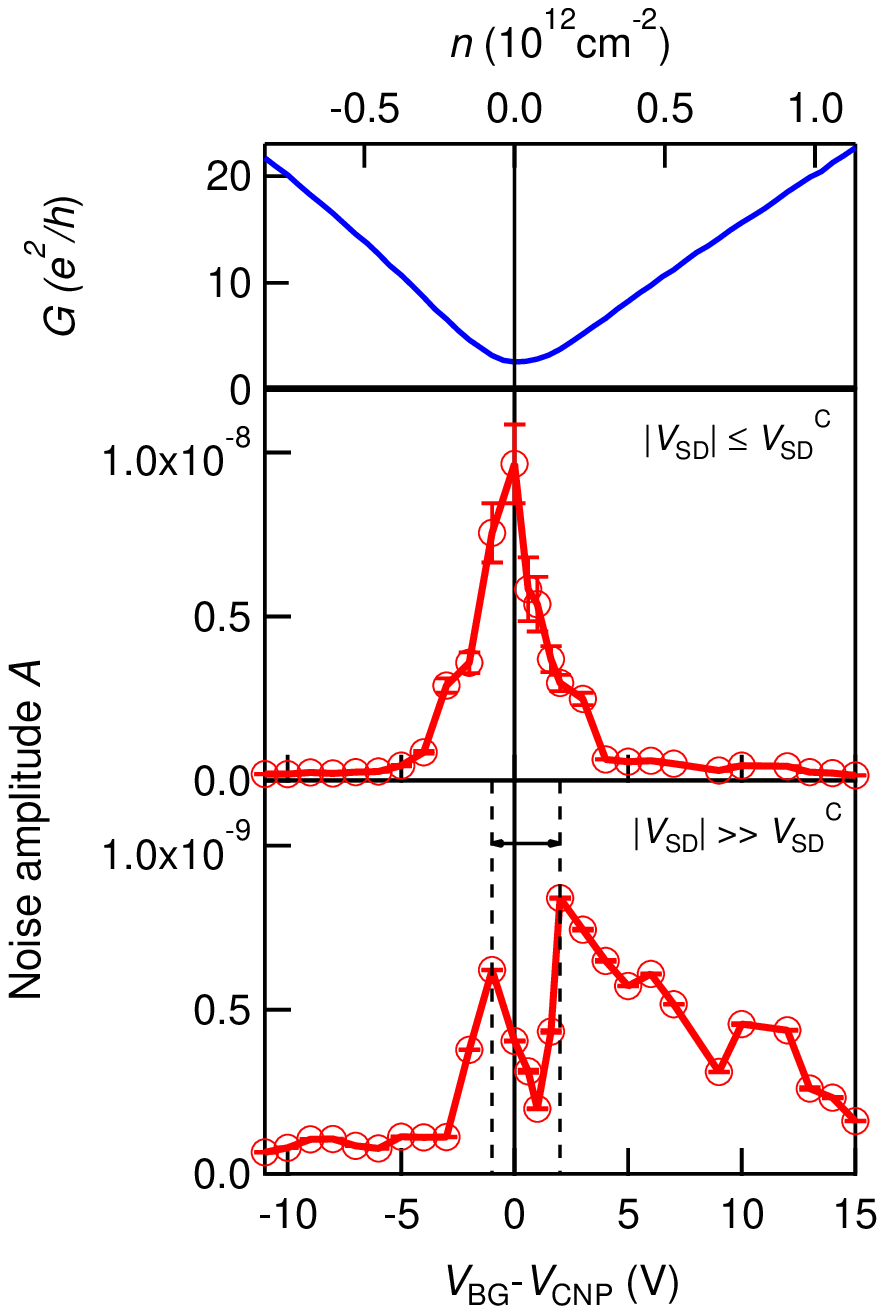}
\end{center}
\caption{Top: Conductance as a function of gate voltage (bottom axis) and carrier density (top axis). The data is the same as shown in Fig.~\ref{Fig1}(a). Middle: Gate voltage dependence of the noise amplitude in low $V_{\mathrm{SD}}$ regime. Bottom: Gate voltage dependence of the noise amplitude in high $V_{\mathrm{SD}}$ regime. The dashed lines indicate the point where the $A$ takes its local maximum. The distance between the two lines corresponds to $2n_\mathrm{max}\sim 2.4\times10^{11}\,\mathrm{cm^{-2}}$. }  
\label{Fig3}
\end{figure}

The numerical fitting is performed using the relation $S_V\cdot$$f=AV_{\mathrm{SD}}^2$ to obtain the dimensionless noise amplitude $A$. In this analysis, we separately deduce the $A$ in the low $V_{\mathrm{SD}}$ regime ($|V_{\mathrm{SD}}|\leq V_{\mathrm{SD}}^\mathrm{C}$) and the high  $V_{\mathrm{SD}}$ regime ($|V_{\mathrm{SD}}|\gg V_{\mathrm{SD}}^\mathrm{C}$) at every gate voltage $V_\mathrm{{BG}}$.
$V_{\mathrm{BG}}$-dependence of $A$ for $|V_{\mathrm{SD}}| \leq V_{\mathrm{SD}}^\mathrm{C}$ and for $|V_{\mathrm{SD}}| \gg V_{\mathrm{SD}}^\mathrm{C}$ are shown in the middle and bottom panels of Fig.~\ref{Fig3}, respectively, together with the conductance in the top panel. The bottom axis is the gate voltage $V_{\mathrm{BG}}- V_{\mathrm{CNP}}$ and the top axis represents the carrier density $n$ which is calculated from the relation $n=C_{\mathrm{BG}} (V_{\mathrm{BG}}-V_{\mathrm{CNP}})/e$ ($C_{\mathrm{BG}}$ is the back gate capacitance per unit area). The deduced noise amplitudes $A$ are of the order of $10^{-8}$ or $10^{-9}$, being consistent with the previous studies~\cite{M1,M3,M4,M6,NM1,NM2,NM3}. 

The behaviors of the noise around CNP in the low $V_{SD}$ region and in the high $V_{SD}$ region are very different between each other. In the former, it is maximum at the CNP and decreases as $|V_{\mathrm{BG}}-V_{\mathrm{CNP}}|$ increases, resulting in the ``$\Lambda$-shape'' gate dependence as is expected from Hooge's relation. In the latter, on the other hand, the noise amplitude is minimum around the CNP, and as the distance from the CNP increases, it increases until  $|V_{\mathrm{BG}}-V_{\mathrm{CNP}}|\sim1\,\mathrm{V}$, and then it decreases again. This ``M-shape'' gate dependence was often reported in previous studies~\cite{M1,M2,M3,M4,M5,M6}. Away from the CNP, the $A$'s are as low as $\ll3\times10^{-10}$ both in low bias and high bias regimes. They are slightly bigger in the electron side ($n>0$) than those in the hole side ($n<0$), which may reflect the difference of the mobilities ($\mu_h>\mu_e$ in the present case).

Here we discuss a possible scenario that could explain the observed behaviors. It is widely accepted that charged impurities on the substrate induce potential fluctuation in the graphene and create spatial charge inhomogeneities (electron-hole puddles~\cite{puddle}). Away from the CNP, the hole (electron) puddles are completely filled by electron (hole) whole of the sample. As a result, the puddles do not largely affect the conduction. Around the CNP, when the electrons (holes) come into the hole (electron) puddles, they need finite activation energy to escape from the regions. Around zero temperature the carriers in the puddles are frozen, or in other words, the puddles are pinned~\cite{Disorder}. In such a situation, the carriers in the puddles cannot contribute to the screening of the scattering centers in the low $V_{\mathrm{SD}}$ regime, and therefore, the device simply behaves as a disordered system with very low carrier density. Here, Hooge's relation is expected to hold as seen in our observation.

Now, what happens to the puddles at high $V_{\mathrm{SD}}$ regime? To explain the observed crossover from the ``$\Lambda$-shape'' to ``M-shape'', we expect a physical mechanism to reduce the effect of the puddles. Thus we would propose that, in the high  $V_{\mathrm{SD}}$ regime, presumably due to the applied electric field across the device and/or due to the injected carriers, the puddles are ``depinned'', and the carriers trapped by those puddles start to move. This assumption could explain the reduction of the $1/f$ noise at the CNP at high  $V_{\mathrm{SD}}$ regime, as these carriers now participate to screen the scattering centers. The slight increase of the conductance around $|V_\mathrm{SD}|=V_\mathrm{SD}^\mathrm{C}$ as seen in Fig.~\ref{Fig2}(b) may reflect this phenomenon. As the distance from the CNP increases, the depining of the puddles, if any, do not affect the noise amplitude, as there are sufficient carriers already. In the previous studies, the measurements were performed in the high  $V_{\mathrm{SD}}$ regime, and only the ``M-shape'' behavior was observed~\cite{M1,M2,M3,M4,M5,M6}. 

To estimate the influence of the puddle effect, we calculate the local carrier density of the puddles. According to \cite{Aself}, the carrier density in the puddles ($n_{\mathrm{puddle}}$) and the density of the charged impurities in the sample ($n_{\mathrm{imp}}$) is calculated from the carrier density dependence of the conductivity to be $|n_{\mathrm{puddle}}|\sim1.2\times10^{11}\,\mathrm{cm^{-2}}$ and $n_{\mathrm{imp}} \sim10\,\times10^{11}\mathrm{cm^{-2}}$ respectively. This $n_{\mathrm{imp}}$ is the typical value for the disordered graphene on $\mathrm{Si/SiO_2}$ with a similar mobility~\cite{Measurementof}. $|n_{\mathrm{puddle}}|\sim1.2\times10^{11}\,\mathrm{cm^{-2}}$ is consistent with the carrier density where the noise amplitude in high bias regime shows local maximum $|n_{\mathrm{max}}|\sim1.2\times10^{11}\,\mathrm{cm^{-2}}$(shown in the bottom panel of Fig.\ref{Fig3}). In addition, it was recently reported that while the noise of the graphene device on $\mathrm{Si/SiO_2}$ showed the ``M-shape'' gate dependence, the one on h-BN showed ``$\Lambda$-shape''~\cite{NM3}. These facts may support our idea that the puddles are responsible for this anomolous noise behavior.

In conclusion, we observed the anomalous behaviors in the $1/f$ noise in graphene; in the low bias voltage regime, the noise shows ``$\Lambda$-shape'' gate dependence, which is consistent with Hooge's relation, while in the high bias regime the noise exhibits well-known ``M-shape'' dependence. These different noise behaviors may result from the effect of the electron-hole puddles and their depining induced by the source-drain bias voltage. Our work shows that the $1/f$ noise study could give us a clue to understand the dynamics of the electronic state around CNP and thus contribute to the realization of low-noise graphene devices.

%\label{}
%\subsection{}
%\subsubsection{}

% If in two-column mode, this environment will change to single-column format so that long equations can be displayed. 
%\begin{widetext}
%$$\mbox{put long equation here}$$
%\end{widetext}

% Figures should be put into the text as floats. 
% Use the graphics or graphicx packages (distributed with LaTeX2e).
% Put the label that you will use with \ref{} command in the braces of the \label{} command.
%
% \begin{figure}
% \end{figure}

% \begin{table}
% \end{tabular}
% \end{table}

\begin{acknowledgments}
This work was partially supported by JSPS KAKENHI (26220711, 15K17680, 25103003, 15H05854, 25107004). K.K. acknowledges the stimulating discussions in the meeting of the Cooperative Research Project of RIEC, Tohoku University.
\end{acknowledgments}

\providecommand{\noopsort}[1]{}\providecommand{\singleletter}[1]{#1}%


\begin{thebibliography}{19}
\expandafter\ifx\csname natexlab\endcsname\relax\def\natexlab#1{#1}\fi
\expandafter\ifx\csname bibnamefont\endcsname\relax
  \def\bibnamefont#1{#1}\fi
\expandafter\ifx\csname bibfnamefont\endcsname\relax
  \def\bibfnamefont#1{#1}\fi
\expandafter\ifx\csname citenamefont\endcsname\relax
  \def\citenamefont#1{#1}\fi
\expandafter\ifx\csname url\endcsname\relax
  \def\url#1{\texttt{#1}}\fi
\expandafter\ifx\csname urlprefix\endcsname\relax\def\urlprefix{URL }\fi
\providecommand{\bibinfo}[2]{#2}
\providecommand{\eprint}[2][]{\url{#2}}

\bibitem[{\citenamefont{Schwierz}(2010)}]{Graphenetransistor}
\bibinfo{author}{\bibfnamefont{F.}~\bibnamefont{Schwierz}},
  \bibinfo{journal}{Nature Nanotech.} \textbf{\bibinfo{volume}{5}},
  \bibinfo{pages}{487} (\bibinfo{year}{2010}).

\bibitem[{\citenamefont{Balandin}(2013)}]{1/fnoisereview}
\bibinfo{author}{\bibfnamefont{A.~A.} \bibnamefont{Balandin}},
  \bibinfo{journal}{Nature. Nanotech.} \textbf{\bibinfo{volume}{8}},
  \bibinfo{pages}{549} (\bibinfo{year}{2013}).

\bibitem[{\citenamefont{Hooge}(1969)}]{Hooge}
\bibinfo{author}{\bibfnamefont{N.~F.} \bibnamefont{Hooge}},
  \bibinfo{journal}{Phys. Lett. A} \textbf{\bibinfo{volume}{29}},
  \bibinfo{pages}{139} (\bibinfo{year}{1969}).

\bibitem[{\citenamefont{Xu et~al.}(2010)\citenamefont{Xu, Carlos M.~Torres,
  Zhang, Liu, Song, Wang, Zhou, Zeng, and Wang}}]{M1}
\bibinfo{author}{\bibfnamefont{G.}~\bibnamefont{Xu}},
  \bibinfo{author}{\bibfnamefont{J.}~\bibnamefont{Carlos M.~Torres}},
  \bibinfo{author}{\bibfnamefont{Y.}~\bibnamefont{Zhang}},
  \bibinfo{author}{\bibfnamefont{F.}~\bibnamefont{Liu}},
  \bibinfo{author}{\bibfnamefont{E.~B.} \bibnamefont{Song}},
  \bibinfo{author}{\bibfnamefont{M.}~\bibnamefont{Wang}},
  \bibinfo{author}{\bibfnamefont{Y.}~\bibnamefont{Zhou}},
  \bibinfo{author}{\bibfnamefont{C.}~\bibnamefont{Zeng}}, \bibnamefont{and}
  \bibinfo{author}{\bibfnamefont{K.~L.} \bibnamefont{Wang}},
  \bibinfo{journal}{Nano Lett.} \textbf{\bibinfo{volume}{10}},
  \bibinfo{pages}{3312} (\bibinfo{year}{2010}).

\bibitem[{\citenamefont{Heller et~al.}(2010)\citenamefont{Heller, Chatoor,
  Mannik, Zevenbergen, Oostinga, Morpurgo, Dekker, and Lemay}}]{M2}
\bibinfo{author}{\bibfnamefont{I.}~\bibnamefont{Heller}},
  \bibinfo{author}{\bibfnamefont{S.}~\bibnamefont{Chatoor}},
  \bibinfo{author}{\bibfnamefont{J.}~\bibnamefont{Mannik}},
  \bibinfo{author}{\bibfnamefont{M.~A.~G.} \bibnamefont{Zevenbergen}},
  \bibinfo{author}{\bibfnamefont{J.~B.} \bibnamefont{Oostinga}},
  \bibinfo{author}{\bibfnamefont{A.~F.} \bibnamefont{Morpurgo}},
  \bibinfo{author}{\bibfnamefont{C.}~\bibnamefont{Dekker}}, \bibnamefont{and}
  \bibinfo{author}{\bibfnamefont{S.~G.} \bibnamefont{Lemay}},
  \bibinfo{journal}{Nano Lett.} \textbf{\bibinfo{volume}{10}},
  \bibinfo{pages}{1563} (\bibinfo{year}{2010}).

\bibitem[{\citenamefont{Zhang et~al.}(2010)\citenamefont{Zhang, Mendez, and
  Du}}]{M3}
\bibinfo{author}{\bibfnamefont{Y.}~\bibnamefont{Zhang}},
  \bibinfo{author}{\bibfnamefont{E.~E.} \bibnamefont{Mendez}},
  \bibnamefont{and} \bibinfo{author}{\bibfnamefont{X.}~\bibnamefont{Du}},
  \bibinfo{journal}{ACS nano} \textbf{\bibinfo{volume}{10}},
  \bibinfo{pages}{1563} (\bibinfo{year}{2010}).

\bibitem[{\citenamefont{Rumyantsev et~al.}(2010)\citenamefont{Rumyantsev, Liu,
  Stillman, Shur, and Balandin}}]{M4}
\bibinfo{author}{\bibfnamefont{S.}~\bibnamefont{Rumyantsev}},
  \bibinfo{author}{\bibfnamefont{G.}~\bibnamefont{Liu}},
  \bibinfo{author}{\bibfnamefont{W.}~\bibnamefont{Stillman}},
  \bibinfo{author}{\bibfnamefont{M.}~\bibnamefont{Shur}}, \bibnamefont{and}
  \bibinfo{author}{\bibfnamefont{A.~A.} \bibnamefont{Balandin}},
  \bibinfo{journal}{Phys. Condens. Matt.} \textbf{\bibinfo{volume}{22}},
  \bibinfo{pages}{395302} (\bibinfo{year}{2010}).

\bibitem[{\citenamefont{Kaverzin et~al.}(2012)\citenamefont{Kaverzin, Mayorov,
  Shytov, and Horsell}}]{M5}
\bibinfo{author}{\bibfnamefont{A.~A.} \bibnamefont{Kaverzin}},
  \bibinfo{author}{\bibfnamefont{A.~S.} \bibnamefont{Mayorov}},
  \bibinfo{author}{\bibfnamefont{A.}~\bibnamefont{Shytov}}, \bibnamefont{and}
  \bibinfo{author}{\bibfnamefont{D.~W.} \bibnamefont{Horsell}},
  \bibinfo{journal}{Phys. Rev. B} \textbf{\bibinfo{volume}{85}},
  \bibinfo{pages}{075435} (\bibinfo{year}{2012}).

\bibitem[{\citenamefont{Stolyarov et~al.}(2015)\citenamefont{Stolyarov, Liu,
  Rumyantsev, Shur, and Balandin}}]{M6}
\bibinfo{author}{\bibfnamefont{M.~A.} \bibnamefont{Stolyarov}},
  \bibinfo{author}{\bibfnamefont{G.}~\bibnamefont{Liu}},
  \bibinfo{author}{\bibfnamefont{S.~L.} \bibnamefont{Rumyantsev}},
  \bibinfo{author}{\bibfnamefont{M.}~\bibnamefont{Shur}}, \bibnamefont{and}
  \bibinfo{author}{\bibfnamefont{A.~A.} \bibnamefont{Balandin}},
  \bibinfo{journal}{Appl. Phys. Lett.} \textbf{\bibinfo{volume}{107}},
  \bibinfo{pages}{023106} (\bibinfo{year}{2015}).

\bibitem[{\citenamefont{Pal et~al.}(2011)\citenamefont{Pal, Ghatak, Kochat,
  Sneha, Sampathkumar, Raghavan, and Ghosh}}]{NM1}
\bibinfo{author}{\bibfnamefont{A.~N.} \bibnamefont{Pal}},
  \bibinfo{author}{\bibfnamefont{S.}~\bibnamefont{Ghatak}},
  \bibinfo{author}{\bibfnamefont{V.}~\bibnamefont{Kochat}},
  \bibinfo{author}{\bibfnamefont{E.~S.} \bibnamefont{Sneha}},
  \bibinfo{author}{\bibfnamefont{A.}~\bibnamefont{Sampathkumar}},
  \bibinfo{author}{\bibfnamefont{S.}~\bibnamefont{Raghavan}}, \bibnamefont{and}
  \bibinfo{author}{\bibfnamefont{A.}~\bibnamefont{Ghosh}},
  \bibinfo{journal}{ACS Nano} \textbf{\bibinfo{volume}{5}},
  \bibinfo{pages}{2075} (\bibinfo{year}{2011}).

\bibitem[{\citenamefont{Liu et~al.}(2009)\citenamefont{Liu, Stillman,
  Rumyantsev, Shao, Shur, and Balandin}}]{NM2}
\bibinfo{author}{\bibfnamefont{G.}~\bibnamefont{Liu}},
  \bibinfo{author}{\bibfnamefont{W.}~\bibnamefont{Stillman}},
  \bibinfo{author}{\bibfnamefont{S.}~\bibnamefont{Rumyantsev}},
  \bibinfo{author}{\bibfnamefont{Q.}~\bibnamefont{Shao}},
  \bibinfo{author}{\bibfnamefont{M.}~\bibnamefont{Shur}}, \bibnamefont{and}
  \bibinfo{author}{\bibfnamefont{A.~A.} \bibnamefont{Balandin}},
  \bibinfo{journal}{Appl. Phys. Lett.} \textbf{\bibinfo{volume}{95}},
  \bibinfo{pages}{033103} (\bibinfo{year}{2009}).

\bibitem[{\citenamefont{Kayyalha and Chen}(2015)}]{NM3}
\bibinfo{author}{\bibfnamefont{M.}~\bibnamefont{Kayyalha}} \bibnamefont{and}
  \bibinfo{author}{\bibfnamefont{Y.~P.} \bibnamefont{Chen}},
  \bibinfo{journal}{Appl. Phys. Lett.} \textbf{\bibinfo{volume}{107}},
  \bibinfo{pages}{113101} (\bibinfo{year}{2015}).

\bibitem[{\citenamefont{Novoselov et~al.}(2004)\citenamefont{Novoselov, Geim,
  Morozov, Jiang, Zhang, Dubonos, Grigorieva, and Firsov}}]{Electricfield}
\bibinfo{author}{\bibfnamefont{K.~S.} \bibnamefont{Novoselov}},
  \bibinfo{author}{\bibfnamefont{A.~K.} \bibnamefont{Geim}},
  \bibinfo{author}{\bibfnamefont{S.~V.} \bibnamefont{Morozov}},
  \bibinfo{author}{\bibfnamefont{D.}~\bibnamefont{Jiang}},
  \bibinfo{author}{\bibfnamefont{Y.}~\bibnamefont{Zhang}},
  \bibinfo{author}{\bibfnamefont{S.~V.} \bibnamefont{Dubonos}},
  \bibinfo{author}{\bibfnamefont{I.~V.} \bibnamefont{Grigorieva}},
  \bibnamefont{and} \bibinfo{author}{\bibfnamefont{A.~A.}
  \bibnamefont{Firsov}}, \bibinfo{journal}{Science}
  \textbf{\bibinfo{volume}{306}}, \bibinfo{pages}{666} (\bibinfo{year}{2004}).

\bibitem[{\citenamefont{Arakawa et~al.}(2015)\citenamefont{Arakawa, Shiogai,
  Ciorga, Utz, Schuh, Kohda, Nitta, Bougeard, Weiss, Ono et~al.}}]{Arakawa}
\bibinfo{author}{\bibfnamefont{T.}~\bibnamefont{Arakawa}},
  \bibinfo{author}{\bibfnamefont{J.}~\bibnamefont{Shiogai}},
  \bibinfo{author}{\bibfnamefont{M.}~\bibnamefont{Ciorga}},
  \bibinfo{author}{\bibfnamefont{M.}~\bibnamefont{Utz}},
  \bibinfo{author}{\bibfnamefont{D.}~\bibnamefont{Schuh}},
  \bibinfo{author}{\bibfnamefont{M.}~\bibnamefont{Kohda}},
  \bibinfo{author}{\bibfnamefont{J.}~\bibnamefont{Nitta}},
  \bibinfo{author}{\bibfnamefont{D.}~\bibnamefont{Bougeard}},
  \bibinfo{author}{\bibfnamefont{D.}~\bibnamefont{Weiss}},
  \bibinfo{author}{\bibfnamefont{T.}~\bibnamefont{Ono}}, \bibnamefont{et~al.},
  \bibinfo{journal}{Phys. Rev. Lett.} \textbf{\bibinfo{volume}{114}},
  \bibinfo{pages}{016601} (\bibinfo{year}{2015}).

\bibitem[{\citenamefont{Matsuo et~al.}(2015)\citenamefont{Matsuo, Takeshita,
  Tanaka, Nakaharai, Tsukagoshi, Moriyama, Ono, and Kobayashi}}]{Matsuo}
\bibinfo{author}{\bibfnamefont{S.}~\bibnamefont{Matsuo}},
  \bibinfo{author}{\bibfnamefont{S.}~\bibnamefont{Takeshita}},
  \bibinfo{author}{\bibfnamefont{T.}~\bibnamefont{Tanaka}},
  \bibinfo{author}{\bibfnamefont{S.}~\bibnamefont{Nakaharai}},
  \bibinfo{author}{\bibfnamefont{K.}~\bibnamefont{Tsukagoshi}},
  \bibinfo{author}{\bibfnamefont{T.}~\bibnamefont{Moriyama}},
  \bibinfo{author}{\bibfnamefont{T.}~\bibnamefont{Ono}}, \bibnamefont{and}
  \bibinfo{author}{\bibfnamefont{K.}~\bibnamefont{Kobayashi}},
  \bibinfo{journal}{Nat. Commun.} \textbf{\bibinfo{volume}{6}},
  \bibinfo{pages}{8066} (\bibinfo{year}{2015}).

\bibitem[{\citenamefont{Martin et~al.}(2008)\citenamefont{Martin, Akenrman,
  Ulbricht, Lohmann, Smet, Klitzing, and Yacoby}}]{puddle}
\bibinfo{author}{\bibfnamefont{J.}~\bibnamefont{Martin}},
  \bibinfo{author}{\bibfnamefont{N.}~\bibnamefont{Akenrman}},
  \bibinfo{author}{\bibfnamefont{G.}~\bibnamefont{Ulbricht}},
  \bibinfo{author}{\bibfnamefont{T.}~\bibnamefont{Lohmann}},
  \bibinfo{author}{\bibfnamefont{J.~H.} \bibnamefont{Smet}},
  \bibinfo{author}{\bibfnamefont{K.~V.} \bibnamefont{Klitzing}},
  \bibnamefont{and} \bibinfo{author}{\bibfnamefont{A.}~\bibnamefont{Yacoby}},
  \bibinfo{journal}{Nat. Phys.} \textbf{\bibinfo{volume}{4}},
  \bibinfo{pages}{144} (\bibinfo{year}{2008}).

\bibitem[{\citenamefont{Li et~al.}(2011)\citenamefont{Li, Hwang, and
  Sarma}}]{Disorder}
\bibinfo{author}{\bibfnamefont{Q.}~\bibnamefont{Li}},
  \bibinfo{author}{\bibfnamefont{E.~H.} \bibnamefont{Hwang}}, \bibnamefont{and}
  \bibinfo{author}{\bibfnamefont{S.~D.} \bibnamefont{Sarma}},
  \bibinfo{journal}{Phys. Rev. B} \textbf{\bibinfo{volume}{84}},
  \bibinfo{pages}{115442} (\bibinfo{year}{2011}).

\bibitem[{\citenamefont{Adam et~al.}(2007)\citenamefont{Adam, Hwang, Galitski,
  and Sarma}}]{Aself}
\bibinfo{author}{\bibfnamefont{S.}~\bibnamefont{Adam}},
  \bibinfo{author}{\bibfnamefont{E.~H.} \bibnamefont{Hwang}},
  \bibinfo{author}{\bibfnamefont{V.~M.} \bibnamefont{Galitski}},
  \bibnamefont{and} \bibinfo{author}{\bibfnamefont{S.~D.} \bibnamefont{Sarma}},
  \bibinfo{journal}{PNAS} \textbf{\bibinfo{volume}{104}}, \bibinfo{pages}{47}
  (\bibinfo{year}{2007}).

\bibitem[{\citenamefont{Tan et~al.}(2007)\citenamefont{Tan, Zhang, Bolotin,
  Zhao, Adam, Hwang, Sarma, Stormer, and Kim}}]{Measurementof}
\bibinfo{author}{\bibfnamefont{Y.~W.} \bibnamefont{Tan}},
  \bibinfo{author}{\bibfnamefont{Y.}~\bibnamefont{Zhang}},
  \bibinfo{author}{\bibfnamefont{K.}~\bibnamefont{Bolotin}},
  \bibinfo{author}{\bibfnamefont{Y.}~\bibnamefont{Zhao}},
  \bibinfo{author}{\bibfnamefont{S.}~\bibnamefont{Adam}},
  \bibinfo{author}{\bibfnamefont{E.~H.} \bibnamefont{Hwang}},
  \bibinfo{author}{\bibfnamefont{S.~D.} \bibnamefont{Sarma}},
  \bibinfo{author}{\bibfnamefont{H.~L.} \bibnamefont{Stormer}},
  \bibnamefont{and} \bibinfo{author}{\bibfnamefont{P.}~\bibnamefont{Kim}},
  \bibinfo{journal}{Phys. Rev. Lett.} \textbf{\bibinfo{volume}{99}},
  \bibinfo{pages}{666} (\bibinfo{year}{2007}).
\end{thebibliography}
\end{document}